\documentclass[aps,floatfix,showpacs,showkeys,preprintnumbers,amsmath,amssymb]{revtex4}
\usepackage[T1]{fontenc}
\usepackage[latin1]{inputenc}
\usepackage{graphicx}% Include figure files
\usepackage{dcolumn}% Align table columns on decimal point
\usepackage{color}

\newcommand{\be}{\begin{equation}}
\newcommand{\ee}{\end{equation}}
\newcommand{\bea}{\begin{eqnarray}}
\newcommand{\eea}{\end{eqnarray}}

\begin{document}

\title{Nuclear response functions in homogeneous matter with
finite range effective interactions}

\author{J. Margueron}
\affiliation{Institut de Physique Nucl\'eaire, Universit\'e Paris Sud F-91406
Orsay CEDEX, France}
\author{J. Navarro}
\affiliation{IFIC (CSIC - Universidad de Valencia), Apdo. 22085,
E-46.071-Valencia, Spain}
\author{Nguyen Van Giai}
\affiliation{Institut de Physique Nucl\'eaire, Universit\'e Paris Sud F-91406
Orsay CEDEX, France}

\date{ \today}

\begin{abstract}
The question of nuclear response functions in a homogeneous medium
is examined. A general method for calculating response functions
in the random phase approximation (RPA) with exchange is
presented. The method is applicable for finite-range nuclear
interactions. Examples are shown in the case of symmetric nuclear
matter described by a Gogny interaction. It is found that the
convergence of the results with respect to the multipole
truncation is quite fast. Various approximation schemes such as
the Landau approximation, or the Landau approximation for the
exchange terms only, are discussed in comparison with the exact
results.
\end{abstract}


\maketitle

\section{Introduction}
Infinite nuclear matter as a homogeneous medium made of interacting
nucleons is not a system that can be experimentally studied in the
laboratory, but it is nevertheless a very useful and broadly used
concept because of its relative simplicity and its connection with
the inner part of atomic nuclei. In the remote environment of our
planet Earth this idealized system can modelize some parts of the
compact stars. It is therefore important to have a microscopic
description of nuclear matter based on nucleon-nucleon interactions.
There are basically two main approaches, either by starting from a
bare two-body force and treating the many-body problem by
Monte-Carlo methods~\cite{wiringa,akmal} or Brueckner-Hartree-Fock
method~\cite{Catania,Machleidt,Malfliet}, or using directly an
effective nucleon-nucleon interaction adjusted to describe the bulk
properties of nuclear matter and finite nuclei in a mean field
approximation. In the latter approach there are two types of
interactions very widely used in a non-relativistic framework,
namely the Skyrme-type forces~\cite{vau72,cha97} and the Gogny-type
forces~\cite{gog75}. In this work we concentrate on the question of
nuclear response functions using finite-range forces like the Gogny
force.

There are many physical issues that require the knowledge of the
response function of the medium to an external probe. Well-known
examples are the electron scattering by nuclei or the propagation
of neutrinos in nuclear matter. In a mean field framework the
response functions must take into account the effects of
long-range correlations by the Random Phase Approximation (RPA)
which is the small amplitude limit of a time-dependent mean field
approach. For contact interactions of the Skyrme type the RPA
response functions have been often studied (see, e.g.,
Ref.~\cite{gar92}). On the other hand, RPA studies of nuclear
matter with Gogny forces are more rare, and they usually involve
some limiting assumption such as the Landau limit, or the small
momentum transfer limit~\cite{gog77}.

It is worthwhile at this point to clarify the terminology used in
the literature. One often uses the short-hand name of RPA for the
{\it ring approximation} of RPA which is obtained when the
particle-hole (p-h) interaction is approximated by its Landau-Migdal
form. Here, the main purpose is to treat exactly the exchange
contributions of the p-h interaction and therefore, we keep the name
RPA for this situation unless otherwise specified. This corresponds
to what is called RPA with exchange (RPAE) in the electron gas physics.

The method for solving the RPA equation with a finite range force
is simple in principle. We show that, with a small number of terms
in the multipole expansions the convergence is fast and
the calculations are relatively easy. We also compare the exact
RPA response functions with various approximations, namely the
Landau-Migdal approximation made on the complete p-h interaction
or on the exchange part of it. The latter approximation keeps the
exact momentum transfer dependence of the direct interaction but
it still has the simplicity of the Landau-Migdal treatment and
therefore, it can be useful for very extensive
studies of particle propagation inside matter.

The outline of the paper is as follows. In Sec.~II we present the
general method for calculating RPA response functions with direct
and exchange p-h interactions. In Sec.~III we discuss the
convergence of the multipole expansion using the Gogny force D1S.
In Sec.~IV we compare the exactly calculated RPA response functions
with approximations of Landau-Migdal type. Concluding remarks are
in Sec.~V.

\section{Formalism}
\label{formalism}

\begin{figure}[h]
\includegraphics[scale=0.25]{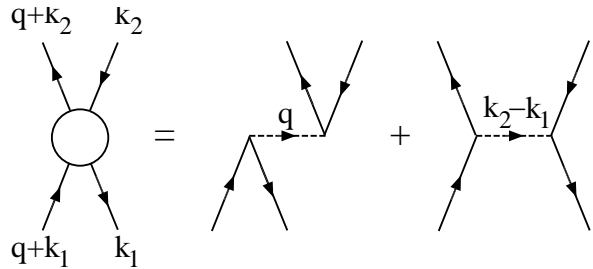}
\caption{Direct and Exchange parts of the ph interaction.}
\label{qk1k2}
\end{figure}

A general two-body interaction in momentum representation depends
at most on 4 momenta. Because of momentum conservation there are
actually 3 independent momenta. For the particle-hole (p-h) case
we choose these independent variables to be the initial (final)
momentum ${\bf k_1}$ (${\bf k_2}$) of the hole
and the external momentum transfer ${\bf q}$. This is illustrated by
Fig.~\ref{qk1k2}. We will denote by
$\alpha = (S,T)$ the spin and isospin p-h channels with $S$=0 (1)
for the non spin-flip (spin-flip) channel, and $T$=0 (1) the
isoscalar (isovector) channel. The matrix element of the general
antisymmetrized p-h interaction can be written as:
\bea
\langle
{\bf q}+{\bf k}_1, {\bf k}_1^{-1} | V | {\bf q}+{\bf k}_2,
{\bf k}_2^{-1} \rangle &=&
\sum_{(\alpha)} V_{ph}^{(\alpha)}({\bf q},{\bf k}_1,{\bf k}_2) P^{(\alpha)} ~,
\label{eq1}
\eea
where the projectors are
$P^{(0,0)}=1/g$, $P^{(1,0)}=(\sigma_1\cdot\sigma_2)/g$,
$P^{(0,1)}=(\tau_1\cdot\tau_2)/g$
and $P^{(1,1)}=(\sigma_1\cdot\sigma_2)(\tau_1\cdot\tau_2)/g$
(the factor $g$=4 is the spin-isospin degeneracy).

For a finite range interaction whose spin-isospin dependence is
described by the usual Wigner, Bartlett, Heisenberg and Majorana
terms, the components $V_{ph}^{(\alpha)}({\bf q},{\bf k}_1,{\bf k}_2)$ 
are:
\bea
V_{ph}^{(\alpha)}({\bf q},{\bf k}_1,{\bf k}_2) &
= & D^{(\alpha)} f(q) + E^{(\alpha)} h({\bf k}_1-{\bf k}_2) +
R^{(\alpha)}~.
\label{eq2}
\eea
The direct terms $D^{(\alpha)}f(q)$ depend only on the modulus $q$ while 
the exchange terms $E^{(\alpha)}h({\bf k}_1-{\bf k}_2)$ depend on
${\bf k}_1-{\bf k}_2$. The last term $R^{(\alpha)}$ accounts for
rearrangement contributions and it must be included if the
starting nucleon-nucleon effective interaction has a density
dependence~\cite{rin80}.

As an example, let us consider as the starting interaction $V$ the
effective Gogny force~\cite{gog75} which is often used for nuclear
matter and nuclear structure studies. It consists of a sum of two
Gaussians having different ranges and spin-isospin dependences
supplemented by a contact term depending on the local density. For
such a force, the functions $f(q)$ and $h({\bf k}_1-{\bf k}_2)$
are:
\bea
f(q) & = & e^{-\frac{1}{4}q^2\mu^2} \nonumber \\
h({\bf k}_1-{\bf k}_2) & = &  e^{-\frac{1}{4}({\bf k_1}-{\bf
k_2})^2\mu^2}~,
\label{eq3}
\eea
where $\mu$ is the range
parameter of the Gaussian form factor. For a Gogny-type force, the
direct and exchange contributions to Eq.~(\ref{eq2}) are obtained
by summing over the two Gaussians. The expressions of $D^{(\alpha)}$
and $E^{(\alpha)}$ in terms of the spin-isospin coefficients and the
range of the Gaussian, as well as the rearrangement term
$R^{(\alpha)}$ are shown in Table~1.

\begin{table}[h]
\centering
\begin{tabular}{|c|c|c|c|c|}
\hline
$(S,T)$ & (0,0) & (0,1) & (1,0) & (1,1) \\
\hline $D^{(S,T)}$ & $\pi \sqrt{\pi} \mu^3 (4W+2B-2H-M )$
            & $\pi \sqrt{\pi} \mu^3 (-2H-M )$
        & $\pi \sqrt{\pi} \mu^3 (2B-M )$
        & $\pi \sqrt{\pi} \mu^3 (-M )$  \\
$E^{(S,T)}$ & $\pi \sqrt{\pi} \mu^3 (-W-2B+2H+4M )$
        & $\pi \sqrt{\pi} \mu^3 (-W-2B )$
        & $\pi \sqrt{\pi} \mu^3 (-W+2H )$
        & $\pi \sqrt{\pi} \mu^3 (-W )$ \\
$R^{(S,T)}$ & $\frac{3}{2} (\gamma+1)(\gamma+2) t_3 \rho^{\gamma}$
        & $- \, (1 + 2 x_3) t_3 \rho^{\gamma}$
        & $- \, (1 - 2 x_3) t_3 \rho^{\gamma}$
        & $- \, t_3 \rho^{\gamma}$ \\
\hline
\end{tabular}
\caption{Direct, exchange and density-dependent components of the
p-h interaction corresponding to a Gogny-type interaction. $W, B,
H, M, \mu, t_3, \gamma$ are interaction parameters, $\rho$ is the
local density.}
\label{table1}
\end{table}

Let us consider for simplicity an infinite nuclear medium at zero
temperature and unpolarized both in spin and isospin spaces. At
mean field level this system is described as an ensemble of
independent nucleons moving in a self-consistent mean field
generated by the starting effective interaction treated in the
Hartree-Fock (HF) approximation. The momentum dependent HF mean
field, or self-energy determines the single-particle spectrum
$\epsilon(k)$ and the Fermi level $\epsilon(k_F)$.

To calculate the response of the medium to an external field it is
convenient to introduce the p-h Green's function, or retarded p-h
propagator $G({\bf q},\omega,{\bf k}_1)$. From now on we choose the
$z$ axis along the direction of ${\bf q}$. In the HF approximation,
the p-h Green's function is the free retarded p-h
propagator~\cite{walecka}: 
\be 
G^{(0)}(q,\omega,{\bf k}_1) =
\frac{\theta(k_F- k_1)- \theta(k_F-|{\bf q}+{\bf k_1}|)} {\omega +
\epsilon(k_1) - \epsilon(\vert{\bf q}+{\bf k}_1)\vert + i \eta}~.
\label{eq4} 
\ee 
It is customary to go beyond the HF mean field approximation and to 
take into account the long-range type of correlations by  resumming a 
class of p-h diagrams. One thus obtains the well-known Random Phase
Approximation (RPA)~\cite{walecka} whose correlated Green's function
$G(q,\omega,{\bf k}_1)$ satisfies the Bethe-Salpeter equation: 
\be
G(q,\omega,{\bf k}_1) = G^{(0)}(q,\omega,{\bf k}_1) +
G^{(0)}(q,\omega,{\bf k}_1) \, \int
\frac{{\rm d}^3k_2}{(2 \pi)^3}
V_{ph}(q,{\bf k}_1,{\bf k}_2) G(q,\omega,{\bf k}_2)~. 
\label{eq5}
\ee 
Finally, the response function $\chi(q,\omega)$ in the infinite
medium is related to the p-h Green's function by: 
\be 
\chi(q,\omega)
= g \int \frac{{\rm d}^3 k_1}{(2 \pi)^3} G(q,\omega,{\bf k}_1)~, 
\label{eq6} 
\ee 
Equations (\ref{eq4}- \ref{eq6}) should be understood for each $(S,T)$ channel.
The Lindhard function $\chi^{(0)}$ is obtained when the  
free p-h propagator $G^{(0)}$ is used in Eq.~(\ref{eq6}). 

If the p-h interaction Eq.~(\ref{eq2}) is treated in some
approximation so as to simplify its $\bf{k}_1 - \bf{k}_2$
dependence, then solving Eq.~(\ref{eq5}) can be made easier. For
instance, neglecting the exchange terms $E^{(\alpha)}$, or
treating them in Landau approximation and keeping
only $l=0$ terms lead to the familiar ring approximation of RPA,
and the response function~(\ref{eq6}) can be expressed in terms of
the Lindhard function $\chi^{(0)}$.
In the past there have been studies of nuclear matter response
functions with Gogny-type interactions under simplifying
assumptions. For instance, RPA response functions in the
long-wavelength limit, i.e., for vanishing momentum transfer $q$
have been calculated in Ref.~\cite{gog77}. Actually, that study was
done in the Landau approximation whereas we are looking for a
complete RPA calculation with finite range interactions.

The method presented in this paper allows one to obtain RPA Green's
functions for all values of $q$ and $\omega$ without further
approximation. This will be useful for studying processes such as
the propagation of particles inside nuclear matter. It is a
cumbersome task to solve directly in the 3-dimensional momentum
space the Bethe-Salpeter equation with the full p-h residual
interaction. The general method of solution proposed here is to
expand the Green's functions and the p-h interaction on a complete
basis of spherical harmonics and to transform Eq.~(\ref{eq5}) into a
set of coupled integral equations on the radial momentum (i.e., the
momentum modulus) variable. The expansion of the unperturbed p-h
Green's function is simple because it has no dependence on the angle
$\phi$ of the vector ${\bf k}_1$: 
\be 
G^{(0)}(q,\omega,{\bf k}_1) =
\sum_L G_L^{(0)}(q, \omega, k_1) Y_{L0}(\hat k_1)~. 
\label{eq7} 
\ee
As for the residual interaction Eq.~(\ref{eq2}) it depends on the
modulus $\vert\bf{k}_1 - \bf{k}_2\vert$ and therefore, its expansion
is: 
\be 
V_{ph}(q,{\bf k}_1, {\bf k}_2) = \sum_{LM} V_L(q,k_1,k_2)
Y_{LM}^*(\hat k_1)Y_{LM}(\hat k_2)~. 
\label{eq8} 
\ee 
Because of the structure of the multipole expansion of $V_{ph}$, the 
expansion of the RPA Green's function is similar to that of $G^{(0)}$: 
\be
G(q,\omega,{\bf k}_1) = \sum_L G_L(q, \omega, k_1) Y_{L0} (\hat
k_1)~. 
\label{eq9} 
\ee 
Indeed, by inserting back this expansion of the propagator into 
Eq.~(\ref{eq5}), a consistent result is obtained. Making use of the 
multipole expansions (\ref{eq7}-\ref{eq9}) one can transform the 
Bethe-Salpeter equation into a set of coupled integral equations for the
multipole components of the RPA Green's function: 
\be 
G_L(q,\omega,k_1) = G^{(0)}_L(q,\omega,k_1) +
\sum_{L'}\int \frac{ k_2^2 dk_2}{(2\pi)^3}F_{LL'}(q,\omega;k_1,k_2)
G_{L'}(q, \omega, k_2)~, 
\label{eq10} 
\ee 
where we have defined 
\be
F_{LL'}(q,\omega;k_1,k_2) = \Big(\sum_{\lambda}
\Big[\frac{(2L+1)(2L'+1)}{4\pi
(2\lambda+1)}\Big]^{\frac{1}{2}}(L0L'0\vert \lambda 0)^2
G^{(0)}_{\lambda}(q,\omega,k_1)\Big) V_{L'}(q,k_1,k_2)~,
\label{eq11} 
\ee 
where $(L_1M_1 L_2M_2 | L_3M_3)$ is a Clebsh-Gordan coefficient. 
The angular momenta $L$ and $L'$ entering the above
expressions are unlimited in principle. We will see in the next
section that in practice the convergence is very fast and very few
terms are necessary to obtain a good accuracy. Finally, the response
function Eq.~(\ref{eq6}) can be expressed as: 
\be 
\chi (q,\omega) =
\frac{\sqrt{4\pi}}{(2\pi)^3} g\int G_{L=0}(q,\omega,k)k^2 dk~,
\label{eq12} 
\ee 
where only the $L=0$ multipole of the RPA Green's function is required.
However, one has to solve the full system of coupled
equations~(\ref{eq10}), since the interaction couples different
multipoles. In practice, a complete calculation implies the choice
of a cut-off value $L_{max}$ for the summations on angular momenta,
and a grid of points in momentum space in order to transform the
integrals into discrete sums. Then, Eq.~(\ref{eq10}) is solved by a matrix 
inversion. For the results presented in the next sections we have chosen 
a grid with a constant number of points. With 100 points, we have obtained 
a good comparison of the free response function with its analytic form, the
Lindhard function. The upper limit is $k_{max}=k_F+q$. Hence, for
$q=k_F/10$, $\Delta k$=0.017 fm$^{-1}$ and $k_{max}$=1.71 fm$^{-1}$
and for $q=k_F$, $\Delta k$=0.031 fm$^{-1}$ and $k_{max}$=3.14
fm$^{-1}$.

\section{Symmetric nuclear matter results}
\label{results}

\begin{figure}[h]
\centering
\includegraphics[scale=0.25]{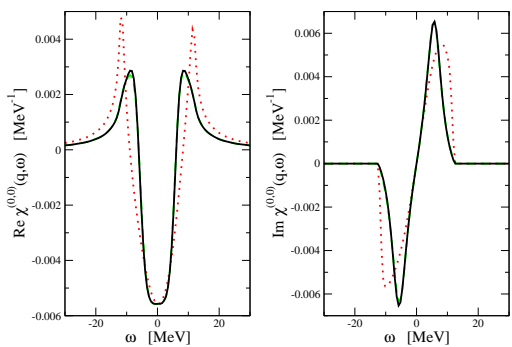}
\vspace{1pt} \caption{Real and imaginary parts of the response function
$\chi^{(0,0)}$ for $q$=27 MeV. Dotted: $L_{max}=0$. Dashed:
$L_{max}=1$. Solid: $L_{max}=2$ (converged).} 
\label{fig1}
\end{figure}

\begin{figure}[h]
\centering
\includegraphics[scale=0.25]{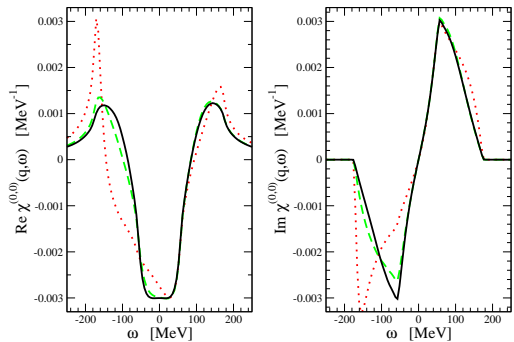}
\vspace{1pt} \caption{Same as Fig.1 for $q$=270 MeV.}
\label{fig2}
\end{figure}

The Gogny force D1S~\cite{gog75} has been chosen to discuss some results 
calculated in symmetric nuclear matter. First, the unperturbed p-h Green's
function $G^{(0)}(q,\omega,{\bf k})$ is calculated using the
Hartree-Fock solution corresponding to D1S. It is just the familiar
Lindhard function and the multipoles $G^{(0)}_L(q,\omega,k)$ can be
easily calculated numerically. Then, an important practical issue is the 
sensitivity of the solution of Eq.~(\ref{eq10}) to the number
of multipoles included in the calculation. To examine this point we
have performed calculations with different values of $L_{max}$. In
Figs.~\ref{fig1}-\ref{fig2} the real and imaginary parts of the RPA response 
function $\chi^{(0,0)}(q,\omega)$ are displayed for several values of
$L_{max}$ and two values of $q$, namely 27 MeV ($\simeq k_F/10$) and
270 MeV ($\simeq k_F$).

It can be seen that for small $q$ the convergence is reached for
$L_{max}$=1 (the solid and dashed curves are indistinguishable in
the figure). Increasing the value of $q$, one sees that
there is still a small difference between $L_{max}$=1 and 2.
The convergence is reached in fact for $L_{max}$=3.

A quantitative measure of the degree of convergence can be provided
by the expected symmetries of the real and imaginary parts of the
response functions. Indeed, the former should be symmetric and the
latter anti-symmetric with respect to $\omega$=0. Let us introduce
for this purpose the following symmetry parameters: 
\bea 
s_r &=&
\int_{\omega \ge 0} {\rm d}\omega \, \vert {\rm Re}\chi(\omega) - 
{\rm Re}\chi(-\omega)\vert  \bigg/ \int_{\omega \ge 0} {\rm d}\omega \, 
\vert {\rm Re}\chi(\omega) + {\rm Re}\chi(-\omega) \vert~, \nonumber \\
s_i &=& 
\int_{\omega \ge 0} {\rm d}\omega \, \vert {\rm Im}\chi(\omega) + 
{\rm Im}\chi(-\omega)\vert \bigg/ \int_{\omega \ge 0} {\rm d}\omega \, 
\vert {\rm Im}\chi(\omega) - {\rm Im}\chi(-\omega)\vert~. 
\label{eq13}
\eea 

The results corresponding to the cases shown in
Figs.~(\ref{fig1}-\ref{fig2}) are indicated in Table~\ref{table2}.
Although the mentioned symmetric behavior is seen in the figures,
the results of Table~\ref{table2} allow one to conclude more quantitatively
that, in the range of momentum transfer up to $k_F$, the symmetry criteria 
are satisfied within 1\% using $L_{max}$=3. In this case, the matrices
involved in the solution of Eq.~(\ref{eq10}) have relatively
moderate sizes (less than 500x500) and the calculations are rather
fast. A similar numerical study for the
response functions in the channels other than ($S=0, T=0$) has been
performed, leading to the same type of convergence as a function of
$L_{max}$. 

\begin{table}[h]
\centering
\begin{tabular}{c|c|c|c|c|c|c|c|}
\cline{3-8}
 \multicolumn{2}{c}{}& \multicolumn{6}{|c|}{$L_{max}$} \\
\cline{2-8}
 & $q$ (MeV) & 0 & 1 & 2 & 3 & 4 & 5\\
\hline
$10^2s_r$ & 27 & 2.10 & 2.27 & 0.42 & 0.06 & - & - \\
      & 270 & 14.7 & 14.0 & 3.75 & 0.94 & 0.56 & 0.52 \\
\hline
$10^2s_i$ & 27 & 3.60 & 2.32 & 0.60 & 0.12 & - & -\\
 & 270 & 14.0 & 13.0 & 3.38 & 0.85 & 0.36 & 0.35 \\
\hline
\end{tabular}
\caption{Symmetry parameters corresponding to the cases shown in
Figs.~\ref{fig1}-\ref{fig2}.}
\label{table2}
\end{table}

To understand this rapid convergence, it is necessary to analyze
in some detail the multipoles of the p-h interaction. To this end, we plot the
dimensionless multipoles 
\be
\tilde{V}_L(q,k_1,k_2) \equiv \frac{2L+1}{4\pi g} N_0 V_L(q,k_1,k_2) 
\label{interaction} 
\ee 
for a fixed value of $q$ in the plane ($k_1/k_F$,$k_2/k_F$) of
hole momenta. In this expression, $N_0=g m^* k_F/(2 \pi^2)$ is the density 
of states, and the factor multiplying $V_L$ fixes the scale: the
value at $q$=0, $k_1=k_2=k_F$ gives the familiar Landau
dimensionless parameter. The factor $4\pi$ comes from the use of
spherical harmonics $Y_{L0}$ instead of Legendre polynomials $P_L$
in the multipole expansions.

\begin{figure}[h]
\center
\includegraphics[scale=0.6]{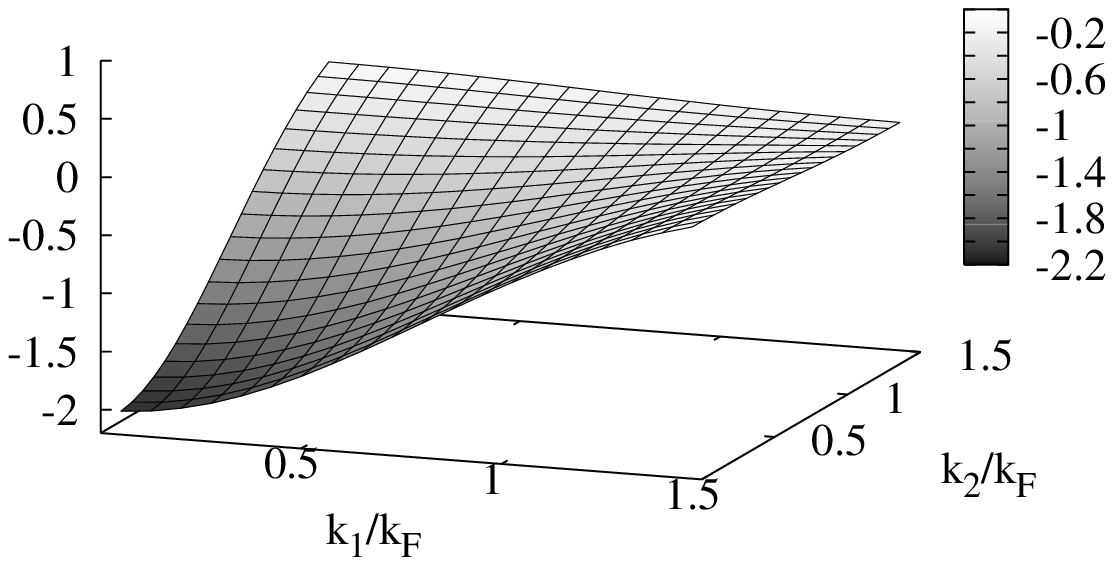}
\includegraphics[scale=0.6]{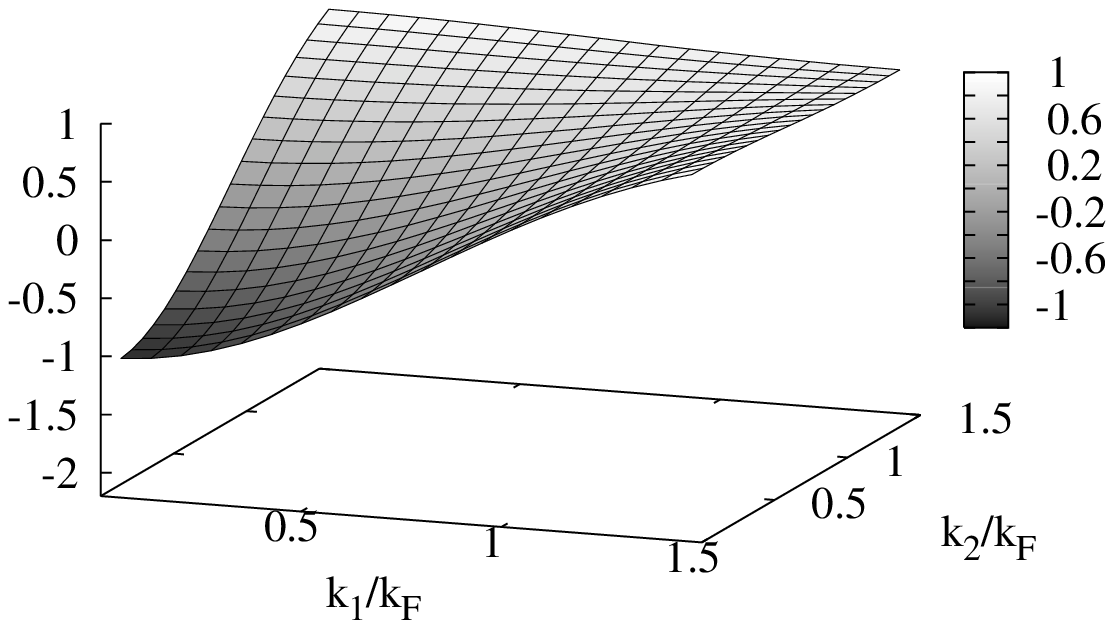}
\caption{The dimensionless monopole $\tilde{V}_0^{(0,0)}(q,k_1,k_2)$ 
as a function of the momenta $k_1/k_F$ and $k_2/k_F$
at $\rho=\rho_0$. The left panel represents the case $q$=0 and the
right one $q$=270 MeV.} 
\label{fig3d1}
\end{figure}

The monopole $L$=0 case is plotted in Fig.~\ref{fig3d1} for
$q$=0 (left panel) and $q$=270 MeV (right panel).
The effect of a finite transferred momentum $q$ only affects the
monopole component of the p-h interaction (see Eq.~\ref{eq2}), and 
it produces an overall translation of the interactions drawn in the figure.
For this specific channel, it induces a repulsion.

\begin{figure}[h]
\center
\includegraphics[scale=0.6]{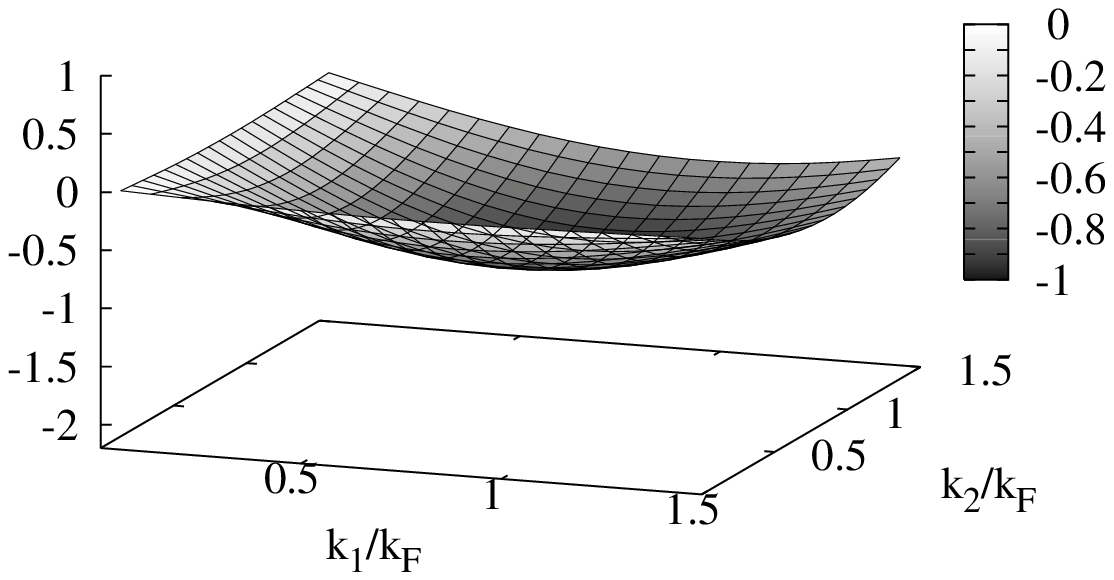}
\includegraphics[scale=0.6]{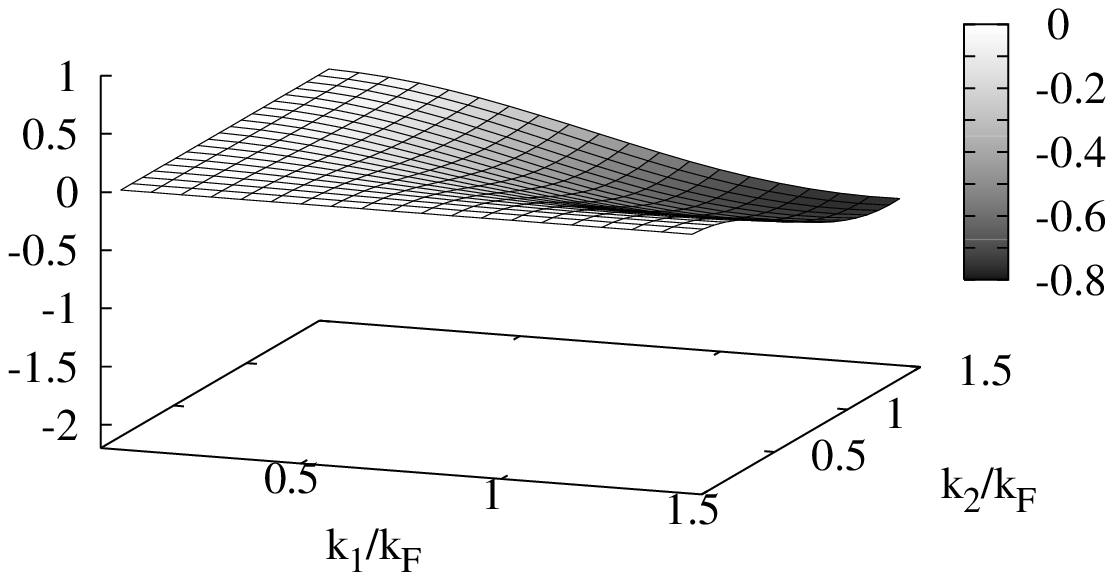}
\caption{The dimensionless multipoles $\tilde{V}_L^{(0,0)}(q,k_1,k_2)$ 
as a function of the momenta $k_1/k_F$ and $k_2/k_F$
at $\rho=\rho_0$. The left panel represents the case $L$=1 and the
right one $L$=2.} 
\label{fig3d2}
\end{figure}

The multipoles $L$=1 and $L$=2 are plotted in Fig.~\ref{fig3d2}, in the 
left and right panels respectively. 
This series of figures for $L$=0,1,2 shows explicitly that the
multipole expansion of the p-h interaction is rapidly converging
after the first few angular momenta.

Inside the ($k_1$,$k_2$) domain considered, the variations of the
multipoles $V_L$ span the range of values (-2:1) for $L$=0, (-1:0) 
for $L$=1, (-0.5:0) for $L$=2, (-0.3:0) for $L$=3
while beyond $L$=4, the multipoles
are practically negligible. In conclusion, the full convergence is
achieved using $L_{max}$=3.

\section{Landau approximations}

Once the convergence of the method has been proved, it is
useful to analyze some approximations employed to obtain the
response function. The Landau-Migdal approximation~\cite{migdal} is
often used in the literature because it simplifies greatly the
calculation of RPA response functions. This approximation was used
in Ref.~\cite{gog77} for the Gogny interaction D1, based on the
solution of the kinetic equations. We should mention that no
analysis of convergence was made. The approximation consists in
assuming that the interacting particle and hole are on the Fermi
surface and that the interaction takes place only in the limit
$q=0$. That is to say that each multipole of the p-h interaction is
replaced by the constant $V_L(0,k_F,k_F)$. The validity of this
assumption can be checked by inspecting Figs.~\ref{fig3d1}-\ref{fig3d2}. 
For the considered interaction and $(S,T)$ channel, the first Landau
dimensionless parameters are
$F_0$=-0.38, $F_1$=-0.91 and $F_2$=-0.33. It can be seen in these
figures that the corresponding multipoles of the p-h interaction are
far from taking a constant value. It is well known that the validity
of the approximation is limited to very small values of $q$, because
in this case the physically
relevant values of $k_1$ and $k_2$ remain around the Fermi momentum.
However, it is worth keeping in mind that in what concerns the
response function the differences between the true and the
approximated p-h interactions are smeared out by the kinetic
constraints of the phase space.

The $q$=0 hypothesis can be easily relaxed. Indeed, Eq.~(\ref{eq2})
shows that only the exchange of the p-h interaction depends on $k_1$
and $k_2$. Thus, we could keep unchanged the direct term
$D^{(\alpha)} f(q)$ of the interaction Eq.~(\ref{eq6}) with its full
$q$-dependence and make the Landau approximation on the exchange
term $E^{(\alpha)} h({\bf k}_1-{\bf k}_2)$ only. We denote this
choice as the Landau Approximation For Exchange Term (LAFET). One
may hope that this procedure will improve the usual Landau
approximation by treating approximately the exchange term only. The
simplicity of the Landau approximation is preserved, the only change
being that the monopole contribution
acquires a $q$-dependence coming from the direct term.

The general method presented in Sec.~\ref{formalism} can be easily
applied to both approximations. In
the appendix~\ref{app_landau}, it is shown that if the p-h
interaction is independent of the hole momenta $k_1$ and $k_2$, the
system of coupled integral equations Eq.~(\ref{eq10}) for the
multipoles $G_L(q,\omega,k)$ can be transformed into a set of
algebraic equations for their integrals $\int dk k^2
G_L(q,\omega,k)$.

\begin{figure}[h]
\centering
\includegraphics[scale=0.3]{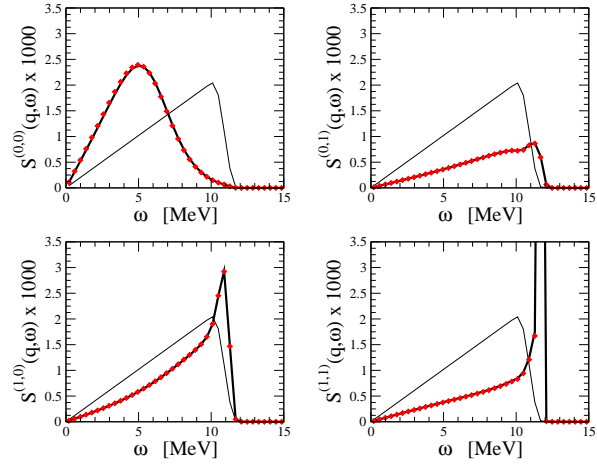}
\vspace{1pt} \caption{Structure function $S^{(S,T)}(q,\omega)$ for
$q$=27 MeV and $L_{max}=3$. Solid thin curve: HF. Dotted: Landau.
Dashed: LAFET. Solid thick line: converged.} \label{figa10}
\end{figure}

\begin{figure}[h]
\centering
\includegraphics[scale=0.3]{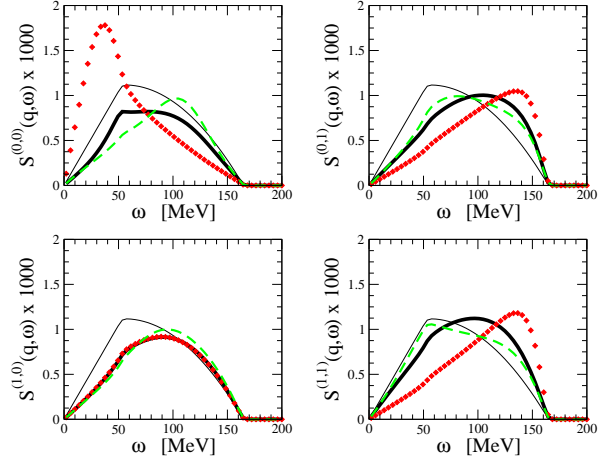}
\vspace{1pt} \caption{Structure function $S^{(S,T)}(q,\omega)$ for
$q$=270 MeV and $L_{max}=3$. Solid thin curve: HF. Dotted: Landau.
Dashed: LAFET. Solid thick line: converged.} \label{figa09}
\end{figure}

We represent on Figs.~\ref{figa10}-\ref{figa09} a comparison of the 
structure function 
\be
S^{(S,T)}(q,\omega)=-\frac{1}{\pi} {\rm Im} \chi^{(S,T)}(q,\omega)
\ee 
extracted from a converged solution (solid thick curve) with
Landau (dotted) and LAFET (dashed) approximations for $q=k_F/10$ and
$q=k_F$. The HF solution (solid thin curve) is also displayed as a reference. 
All the curves shown correspond to $L_{max}$=3, to guarantee the 
convergence of the solution as shown in Sec.~\ref{results}. As
expected, the Landau approximation is a good one for small transferred 
momentum $q$ and all $(S,T)$ channels. However, for the highest
transferred momentum considered in 
this paper, the validity of the Landau approximation and LAFET
response functions is very $(S,T)$ dependent. As a rule, LAFET
induces a response closer to the exact one. Hence, LAFET could be
considered as a very simple extension of the Landau approximation
which allows one to
evaluate processes which involve finite transferred momenta.

There are situations where it is
necessary to calculate accurately response functions of a nuclear
medium to an external probe, as we already mentioned in the
introduction. When the transferred momenta are not small compared to
the Fermi momentum, one must perform the full calculations with the
method presented in Sec.~II. Alternatively, it is possible to use
the LAFET method, which is very simple and efficient if extensive
calculations are needed, and improves the Landau approximation.

\section{Conclusion}

The main purpose of this work is to
present a general method for obtaining nuclear response functions in
an infinite medium within a Hartree-Fock-RPA framework. Starting
with finite range effective interactions like the Gogny interaction,
approximate methods using the Landau approximation are available in
the literature, but surprisingly not much beyond this approximation
can be found. The method proposed here simply consists in expanding
the Bethe-Salpeter equation onto a spherical harmonics basis and
therefore, the calculations can be carried out in principle up to
any degree of accuracy if one includes a sufficient number of
partial waves. In practice, the case study that we have discussed in
this work shows that the convergence is very fast and that the
number of multipoles needed is very small. This result holds not
only for small values of momentum transfer but even at values in the
range of the Fermi momentum. The fast convergence is related to the
properties of the effective p-h interaction.  

This general method of solving the
Bethe-Salpeter equation suggests also an approximation scheme beyond
the standard Landau approximation, the LAFET scheme where the full
$q$-dependence is kept in the direct p-h interaction and the Landau
approximation is done only on the exchange p-h interaction. The
standard Landau approximation and LAFET are compared with the exact
response functions and it is shown that the LAFET results show an
improved agreement with the exact results. This approximation can be
useful for extensive calculations when the numerical effort required
by exact calculations becomes heavy.

\subsection*{Acknowledgments}

This work is supported in part by the grant FIS2004-0912 (MEC,
Spain) and by the IN2P3(France)-CICYT(Spain) exchange program. We
thank P. Schuck for bringing to our attention the work quoted in
Ref.~\cite{gog77}.

\appendix

\section{Landau approximations}
\label{app_landau}

Let us show that if the p-h interaction is independent of the hole
momenta $\bf{k_1}$ and $\bf{k_2}$, the system of coupled integral
equations Eq.~(\ref{eq10}) for the multipoles $G_L(q,\omega,k)$ can
be transformed into a set of algebraic equations for the following
quantities 
\be 
R_L(q,\omega)=g\frac{\sqrt{4\pi}}{(2\pi)^3}\int dk
k^2 G_L(q,\omega,k)~. 
\label{eq50} 
\ee 
The factors before the
integral are chosen such that the response function is given by
$R_{L=0}$.

Let us consider the multipoles of the p-h interaction,
Eq.~(\ref{eq8}), in the particular case where the hole momenta
$\bf{k_1}$ and $\bf{k_2}$ lie on the Fermi surface. For this
specific interaction, the functions $F_{LL'}$ entering
Eq.~(\ref{eq11}) no longer depend on $k_2$. Since all the momentum
dependence is now contained in the p-h propagators, $G_L$ and
$G_L^{(0)}$, integrating over $k_1$ the system Eq.~(\ref{eq10}) can
be transformed into the set of equations: 
\be
R_L(q,\omega)=R_L^{(0)}(q,\omega)+\sum_{LL'}A_{LL'}(q,\omega)
R_{L'}(q,\omega)~, 
\label{eq52} 
\ee 
where $R_L^{(0)}$ is defined as in
Eq.~(\ref{eq50}) for the free propagator and 
\be 
A_{LL'}(q,\omega) =
\frac{{\tilde V}_{L'}(q,k_F,k_F)}{N_0}
\sum_{\lambda}\Big[\frac{(2L+1)}{(2L'+1)(2\lambda+1)}\Big]^{\frac{1}{2}}
(L0L'0\vert \lambda 0)^2 R^{(0)}_{\lambda}(q,\omega)~. 
\label{eq53}
\ee 
It is worth noting that the Landau parameters are given by
$f_L\equiv {\tilde V}_L(q=0,k_F,k_F)/N_0$.

As an example, we give here the response function in the Landau
approximation with $L_{max}$=2: 
\be 
\chi =
\frac{\chi^{(0)}}{1-W\chi^{(0)}}~, 
\label{eq16} 
\ee 
where 
\be 
W = f_0
- \frac{1}{2} f_2 + \nu^2 \frac{ f_1 + ( \frac{27}{8} \nu^2 +
\frac{3}{10} F_1 ) f_2} { \left( 1+\frac{1}{3}F_1 \right) \left( 1 +
\frac{1}{8} [-9\nu^2+\frac{12}{5}]F_2 \right)} 
\ee 
plays the role of the induced p-h interaction. In this expression,
we have defined $\nu = \omega m^*/(q k_F)$, and $F_i=f_i N_0$ are 
the dimensionless Landau parameters.

\end{document}